\documentstyle[eqsecnum,prd,aps,preprint]{revtex}
\tighten
\begin{document}
\draft

\def\beq{\begin{equation}}
\def\eeq{\end{equation}}
\def\beqa{\begin{eqnarray}}
\def\eeqa{\end{eqnarray}}
\def\ben{\begin{enumerate}}
\def\een{\end{enumerate}}
\def\bed{\begin{description}}
\def\eed{\end{description}}
\def\vphi{\varphi}
\def\p{\partial}
\def\a{\alpha}
\def\b{\beta}
\def\g{\gamma}
\def\k{\kappa}
\def\d{\delta}
\def\arrow{\rightarrow}
\def\z{\zeta}
\def\C{{\cal C}}
\def\O{\Omega}
\newcommand{\simg}{\gtrsim}
\newcommand{\siml}{\lesssim}
\def\lkk{\left[}
\def\rkk{\right]}
\def\lmk{\left(}
\def\rmk{\right)}
\def\lnk{\left\{}
\def\rnk{\right\}}
\preprint{
    YITP-97-2, 
    KUNS-1427,
gr-qc/9708030
}

\title{
Imprints of the Metrically-coupled Dilaton on Density Perturbations in
Inflationary Cosmology
}


\author{
Takeshi Chiba$^1$\footnote{
e-mail: chiba@yukawa.kyoto-u.ac.jp; current address: Department of
Physics, University of Tokyo, Tokyo 113-0033, Japan.}, 
Naoshi Sugiyama$^2$ and 
Jun'ichi Yokoyama$^1$
}
\address{
$^1$Yukawa Institute for Theoretical Physics, Kyoto University,
Kyoto 606-8502, Japan}
\address{
$^2$Department of Physics, Kyoto University, Kyoto 606-8502, Japan}
\date{\today}
\maketitle

\begin{abstract}
  Spectra of density perturbations produced during chaotic inflation 
  are calculated, taking both adiabatic and isocurvature modes into
  account in a class of scalar-tensor theories of gravity in  which
  the dilaton is metrically coupled. 
  Comparing the predicted spectrum of the cosmic
  microwave background radiation anisotropies 
  with the one observed by the COBE-DMR we calculate
  constraints on the parameters of these theories, which turn out to
  be stronger by an order-of-magnitude than those
  obtained from post-Newtonian experiments.

\end{abstract}

keywards: inflationary cosmology; scalar-tensor theory of gravity;
linear perturbation theory

\section{INTRODUCTION}

Inflationary cosmology gives a natural explanation for the horizon,
flatness, homogeneity, and monopole problems\cite{inflation}.  In its
simplest form, inflation with a single scalar field predicts an
approximately scale-invariant spectrum of Gaussian adiabatic density
perturbations\cite{dense}. If other fields are present, however, 
primordially isocurvature perturbations may also be generated to leave an
imprint on the universe today \cite{iso,ps}. 

Scalar-tensor theories of gravitation have recently been 
received a renewed interest. 
The main reason is that  the 
low-energy effective action of superstring theories\cite{callan}
generally involves a dilaton coupled to the Ricci curvature. 
Moreover, as pointed out by Damour and Esposite-Farese\cite{de}
 in the study 
of neutron star models, a wide class of scalar-tensor theories not
only pass the present weak-field gravitational tests but also 
exhibit nonperturbative strong-field deviations away from general 
relativity. Detectability of scalar gravitational waves 
by laser interferometric gravitational wave
observatories(LIGO)
has been explored in \cite{snn,hcnn}. 

Cosmological consequences of such theories, however, have not been fully
examined.  Recently, one of us  studied the generality of chaotic
inflation in scalar-tensor theories in detail and found that the onset
of inflation could be greatly affected by the presence of the dilaton
in some cases\cite{chiba}.  According to the attractor mechanism of
scalar-tensor theories\cite{dn}, which  works only for those
theories in which the derivative of the Brans-Dicke coupling is
positive, even those theories that are quite different from general
relativity in the early stage of the universe approach general
relativity during the inflationary stage and the matter-dominated stage
of the universe.
Even if this is the case, the deviations in the earlier stage may be
imprinted in the spectrum of the cosmic microwave background radiation
(CMB).  We may constrain model parameters using the CMB anisotropy as a
probe of the early universe.  In fact, because of the presence of two
scalar fields, {\it i.e.}, the inflaton and the dilaton, density perturbations
have both adiabatic and isocurvature modes and hence the spectrum may
be changed.

In this paper, we study formation and
evolution of density perturbations during
and after chaotic inflation paying attention to the role of the
isocurvature mode in a class of 
scalar-tensor theories in which the dilaton is metrically coupled.
This paper originates from the analysis by
Starobinsky and Yokoyama\cite{sy}, who calculated the spectrum of
density perturbations from inflation in Brans-Dicke theory.  
The scalar-tensor theories we
study include those with negative $\beta$ (the derivative of the
Brans-Dicke coupling) which have not been studied in the literature
\cite{bw} in which the numerical analysis is limited to positive $\beta$. 
These theories exhibit a significant deviation from general relativity 
in the strong field region\cite{de}.

The plan of this paper is as follows: In \S 2, basic equations of
background fields are given. Linear perturbation equations are given and
solutions to these equations are derived in \S 3.  We study the
evolution of isocurvature perturbations after inflation in \S 4. In
\S 5, the spectrum of density perturbations are calculated and compared
with the COBE observation to constrain the parameters of the theory. 
Finally  \S 6 is devoted to summary.

\section{background field equations}

\subsection{scalar-tensor theories of gravitation}

We consider the inflaton, $\sigma$, coupled to scalar-tensor theories of
gravity. The action is 
\beq
S=\int d^4x\sqrt{-\bar{g}}\lkk{1\over 16\pi}\lmk
\psi \bar{\cal R}-{\omega(\psi)\over
\psi}\bar{g}^{ab}\psi_{,a}\psi_{,b}\rmk
-{1\over 2}\bar{g}^{ab}\sigma_{,a}\sigma_{,b}
-V(\sigma)\rkk ,
\eeq
where $\psi$ is the massless 
Brans-Dicke dilaton,  $\bar{\cal R}$ is 
the scalar curvature, $\omega(\psi)$ is a dimensionless coupling
parameter and $V(\sigma)$ is the potential of the inflaton.
Let us consider a conformal transformation,
\beq
\bar{g}_{ab}=\frac{1}{\psi G}g_{ab} \equiv e^{2a(\psi)}g_{ab},
\eeq
with $\psi=G^{-1}$ or $a(\psi)=0$ today.
Then the action can be written as
\beq
S=\int d^4x\sqrt{-g}\lkk{1\over 2\kappa^2}{\cal R}-
{1\over 2}g^{ab}\vphi_{,a}\vphi_{,b}
-{1\over 2}e^{2a}g^{ab}\sigma_{,a}\sigma_{,b}-e^{4a}V(\sigma)\rkk ,
\label{eaction}
\eeq
where $\kappa^2 \equiv 8\pi G$ and $\vphi$ is defined by 
\beq
\vphi \equiv \frac{1}{\kappa}\int \frac{d\psi}{\psi}
\sqrt{\omega(\psi)+\frac{3}{2}}.
\eeq
We also define
\beq
\alpha(\vphi)\equiv \frac{1}{\kappa}\frac{da}{d\vphi},~~~~~
\beta(\vphi)\equiv \frac{1}{\kappa}\frac{d\alpha}{d\vphi},
\eeq 
which are related with the PPN parameters ($\gamma_{Edd},\beta_{Edd}$) 
and are given as
\beq
|\gamma_{Edd} -1| ={4\alpha(\vphi)^2\over
1+2\alpha(\vphi)^2}\bigg{|}_0 ,~~|\beta_{Edd} -1|=
{2\alpha(\vphi)^2|\beta(\vphi)|\over 1+2\alpha(\vphi)^2}\bigg{|}_0,
\label{ppn}\eeq
where the subscript $0$ indicates the present value. 
Here the Brans-Dicke theory corresponds to $\beta(\vphi)=0$ and the general
relativity to $\alpha(\vphi)=0$. 
The recent analyses of the experimental data yield\cite{Will}
\beq 
|\gamma_{Edd} -1| < 2\times
10^{-3},~~|4\beta_{Edd} -\gamma_{Edd} -3| < 1\times 10^{-3} .
\eeq
Combining them yields the following constraint on $\beta_{Edd}$
\beq
|\beta_{Edd} -1| < 6\times 10^{-4}.
\label{ppn2}
\eeq

\subsection{background equations}

We consider the spatially flat FRW spacetime as a background metric
\beq
ds^2=-dt^2+R(t)^2d{\bf x}^2.
\eeq
The equations of motion derived from the action Eq.(\ref{eaction}) are
\beqa
\lmk{\dot{R}\over R}\rmk^2
 \equiv H^2 &=&{\kappa^2\over 6}\lkk\dot{\vphi}^2+e^{2a}
\dot{\sigma}^2+2e^{4a}V(\sigma)\rkk,\label{back1}\\
\ddot{\vphi}+3H\dot{\vphi}&=&\kappa
\alpha\lkk e^{2a}\dot{\sigma}^2-4e^{4a}V(\sigma)\rkk,\label{back2}\\
\ddot{\sigma}+3H\dot{\sigma}&=&-e^{2a}V'(\sigma)-2\kappa\alpha\dot{\vphi}
\dot{\sigma},\label{back3}\\
\dot{H}&=&-{\kappa^2\over
2}(\dot{\vphi}^2+e^{2a}\dot{\sigma}^2), \label{back4}
\eeqa
where $\dot{}$ and $'$ denote time and $\sigma$ derivatives, respectively.

The conditions for slow-roll inflation
($|\ddot{\sigma}|\siml |3H\dot{\sigma}|, |\ddot{\vphi}|\siml
|3H\dot{\vphi}|$) are written as \cite{sy,bw,barrow}, 
\beqa
\max\lmk{1\over 2}\dot{\vphi}^2,{1\over 2}e^{2a}\dot{\sigma}^2\rmk &\siml
&e^{4a}V(\sigma),\label{con1}\\
e^{-2a}V'(\sigma)^2 & \siml &6\kappa^2 
V(\sigma)^2,\label{con2}\\
e^{-2a}V''(\sigma) & \siml & 3\kappa^2 V(\sigma),\label{con3}\\
8\alpha(\vphi)^2&\siml & 3,\label{con4}\\
|16\alpha(\vphi)^2+4\beta(\vphi)|& \siml &3.\label{con5}
\eeqa
When the above inequalities apply, 
field equations are simplified to
\beqa
3H^2 &\simeq & \kappa^2e^{4a}V(\sigma),\label{hams}\\
3H\dot{\vphi} &\simeq & -4\alpha\kappa e^{4a}V(\sigma),\label{vphis}\\
3H\dot{\sigma}& \simeq &-e^{2a}V'(\sigma).\label{sigmas}
\eeqa

\section{linear perturbations}
\subsection{basic equations}

Now we turn to linear perturbations with a perturbed metric in the
longitudinal gauge
\beq
ds^2=-(1+2\Phi)dt^2+R^2(1-2\Psi)d{\bf  x}^2.
\eeq
From the perturbed Einstein equations, each Fourier mode\footnote{
Our convention of the Fourier transformation is 
$Y({\bf k})=(2\pi)^{-3/2}\int d^3x e^{-i{\bf kx}} Y({\bf x})$ for each 
variable $Y$.} 
satisfies the following equations of motion to first 
order\cite{sy,bw},
\beqa
\Phi&=&\Psi,\\
\ddot{\delta\vphi}+3H\dot{\delta\vphi}+{k^2\over R^2}\delta \vphi
&+&4e^{4a}\kappa^2V(\sigma)(4\alpha^2+\beta)\delta \vphi-
e^{2a}\kappa^2\dot{\sigma}^2(2\alpha^2+\beta)\delta \vphi
\nonumber\\
&=&
4\dot{\Phi}\dot{\vphi}+2e^{2a}\alpha\kappa\dot{\sigma}\dot{\delta\sigma}
-4e^{4a}\alpha\kappa(V'(\sigma)\delta\sigma+2V(\sigma)\Phi),\label{vphi}\\
\ddot{\delta\sigma}+3H\dot{\delta\sigma}+{k^2\over R^2}\delta \sigma 
&+&e^{2a}V''(\sigma)\delta\sigma=-2e^{2a}V'(\sigma)(\Phi+\alpha\kappa
\delta\vphi
)\nonumber\\
&+&\dot{\sigma}(4\dot{\Phi}-2\alpha\kappa\dot{\delta\vphi}-2\beta
\kappa^2\dot{\vphi}\delta\vphi)-2\alpha\kappa\dot{\vphi}\dot{\delta\sigma},
\label{sigma}\\
\ddot{\Phi}+4H\dot{\Phi}+(\dot{H}+3H^2)\Phi&=&{\kappa^2\over 2}\lkk
\dot{\vphi}\dot{\delta\vphi}+e^{2a}\dot{\sigma}\dot{\delta\sigma}
-e^{4a}V'(\sigma)\delta\sigma
+(e^{2a}\dot{\sigma}^2-4e^{4a}V(\sigma))\alpha\kappa\delta\vphi\rkk ,\label{Phi}
\eeqa
together with the Hamiltonian and momentum constraints
\beqa
3H\dot{\Phi}+(\dot{H}+3H^2)\Phi+{k^2\over R^2}\Phi&=&
-{\kappa^2\over 2}\lkk\dot{\vphi}\dot{\delta\vphi}+e^{2a}\dot{\sigma}
\dot{\delta\sigma}
+e^{4a}V'(\sigma)\delta\sigma+(e^{2a}\dot{\sigma}^2+
4e^{4a}V(\sigma))\alpha\kappa\delta\vphi\rkk ,\label{hamp}\\
\dot{\Phi}+H\Phi&=&{\kappa^2\over
2}(\dot{\vphi}\delta\vphi+e^{2a}\dot{\sigma}\delta\sigma).\label{momp}
\eeqa

\subsection{long wavelength perturbations}

Since what we need are the non-decreasing adiabatic and isocurvature
modes on large scale $k \ll RH$, which turn out to be weakly
time-dependent as will be seen in the final result \cite{ps}, we may
consistently neglect $\dot{\Phi}$ and those terms containing two time
derivatives.  Then eqs.\ (\ref{vphi}),
(\ref{sigma}), and (\ref{momp})  are  simplified to
\beqa
\Phi &\simeq& -2\alpha\kappa\delta\vphi -{1\over 2}{V'\over V}\delta\sigma,\\
3H\dot{\delta\vphi} &\simeq& -4\beta \kappa^2 e^{4a}V\delta\vphi,\\
3H\dot{\delta\sigma} &\simeq& -e^{2a}\lmk {V'\over
V}\rmk 'V\delta\sigma +2\alpha\kappa e^{2a}V'\delta\vphi.
\eeqa
Following \cite{sy}, we find 
 the last two equations can be integrated to give
\beqa
\delta\vphi &\simeq& {4\alpha\over \kappa}Q_1 ,\\
\delta\sigma &\simeq& {1\over \kappa^2}{V'\over V}(e^{-2a}Q_1+Q_2),
\eeqa
where use has been made of Eqs.(\ref{hams}), (\ref{vphis}) and
(\ref{sigmas}). 
We then find  
\beq
\Phi \simeq -8\alpha^2Q_1-{1\over 2\kappa^2}\lmk {V'\over
V}\rmk ^2(e^{-2a}Q_1+Q_2),
\eeq
where $Q_1$ and $Q_2$ are constants of integration.

\subsection{adiabatic  and isocurvature modes}

In order to clarify the physical meaning of the above solutions, we
should divide them into adiabatic and isocurvature 
modes \cite{sy}\cite{ps}. 
The primordially isocurvature mode is characterized by its vanishingly 
small contribution to the curvature perturbation initially, while the growing 
adiabatic mode can be described by the following expressions in the
long wavelength limit $k \ll RH$ (see
Appendix A for a review of its derivation). 
\beqa
\Phi_{ad} &=& C_1\lkk 1-{H\over R}\int R(t')dt' \rkk  \simeq 
-C_1{\dot{H}\over H^2},\label{ad1}\\
{\delta\vphi _{ad}\over \dot{\vphi}} &=& 
{\delta\sigma _{ad}\over \dot{\sigma}} 
= {C_1\over R}\int R(t')dt' \simeq {C_1\over H},\label{ad2}
\eeqa
as a solution of a second-order differential equation Eq.(\ref{adiab}) 
without the source term. 
Here $C_1$ is a constant and the latter approximate equality in each
expression is derived by integration by parts during 
the inflationary stage. 

As discussed by Starobinsky and Yokoyama \cite{sy} when the conditions
\beqa
  8\alpha^2 &\ll& \frac{1}{2\kappa^2}\lmk\frac{V'}{V}\rmk^2e^{-2a},
\label{cond1}\\
 e^{-2a}&\simeq & 1 
\label{cond2}
\eeqa
hold, one can discriminate between adiabatic and
isocurvature modes in the final results by defining new constants 
$C_1$ and $C_3$ as $C_1 \equiv -Q_1-Q_2$ and $C_3\equiv -Q_2$.
Using Eqs.(\ref{hams}), (\ref{vphis}) and (\ref{sigmas})
we find
\beqa
\Phi &\simeq& -C_1{\dot{H}\over H^2}+C_3\lkk -8\alpha^2+{1\over 2\kappa^2}
\lmk {V'\over V}\rmk ^2
(1-e^{-2a})\rkk ,\label{c3}\\
{\delta\vphi \over \dot{\vphi}}&\simeq& {C_1\over H}-{C_3\over
H},\\
{\delta\sigma \over \dot{\sigma}} &\simeq& {C_1\over H} - {C_3\over
H}(1-e^{2a}).
\eeqa
In the above expressions, terms in proportion to $C_1$ and $C_3$
represent adiabatic and isocurvature modes, respectively if the above
conditions hold. In more general scalar-tensor
theories (especially in positive $\beta$ theories) 
Eq.(\ref{cond1}) and Eq.(\ref{cond2}) may not hold and consequently 
terms in proportion to 
$C_3$  contain the adiabatic mode partially. 
Then these terms might be better referred to as the non-adiabatic
mode. But this is a matter of terminology that does not affect our
final result and we keep using the same expression in this case, too. 

\subsection{quantum fluctuations}
We shall next determine the constants $C_1$ and $C_3$ from amplitudes
of quantum fluctuations of the scalar fields generated during the 
inflationary 
stage. Due to the inequalities (\ref{con1}),(\ref{con2}) and (\ref{con3}),
 Eqs.(\ref{vphi}) and (\ref{sigma}) can be approximated by
the equation of motion of a free massless scalar field during inflation for 
$k\geq RH$. The standard quantization gives the well-known result,
that is, the Fourier components of the fields can be represented in
the form
\beq
\delta\vphi({\bf k})={H(t_k)\over \sqrt{2k^3}}\epsilon_{\vphi}({\bf k}),
~\delta\sigma({\bf k})={H(t_k)\over \sqrt{2k^3}}e^{-a}\epsilon_{\sigma}
({\bf k}),
\label{init}\eeq
where $t_k$ is the time when $k$-mode leaves the Hubble horizon during 
inflation. 
Here $\epsilon_{\vphi}(k)$ and $\epsilon_{\sigma}(k)$ are classical
random Gaussian variables with the following averages:
\beq
\langle\epsilon_{\vphi}({\bf k})\rangle=
\langle\epsilon_{\sigma}({\bf k})\rangle=0,~ 
\langle\epsilon_{i}({\bf k})\epsilon_{j}({\bf k'})\rangle=
\delta_{ij}\delta^{(3)}({\bf k}-{\bf k'}),~
i,j=\vphi,\sigma .
\eeq
We thus find 
\beqa
C_1 &=&\lkk e^{-2a}H{\delta\sigma \over \dot{\sigma}}+
(1-e^{-2a})H{\delta\vphi \over \dot{\vphi}}
\rkk _{t_k}
={H^2(t_k)\over \sqrt{2k^3}}\lkk{e^{-3a}\over
\dot{\sigma}}\epsilon_{\sigma}({\bf k})
+{1-e^{-2a}\over
\dot{\vphi}}\epsilon_{\vphi}({\bf k})\rkk _{t_k},\\
C_3 &=&\lkk e^{-2a} H\lmk  {\delta\sigma 
\over \dot{\sigma}}- {\delta\vphi \over \dot{\vphi}} \rmk \rkk _{t_k}=
{H^2(t_k)\over \sqrt{2k^3}}\lkk {e^{-3a}\over
\dot{\sigma}}\epsilon_{\sigma}({\bf k})-{e^{-2a}\over
\dot{\vphi}}\epsilon_{\vphi}({\bf k})
\rkk _{t_k}.
\eeqa
Note that in Ref.\cite{bw}, they ignored the entropy perturbations 
at and after the end of inflation which corresponds to setting $C_3=0$
when they calculated the spectral index although they originally took
into account these quantities in their derivation.  This cannot be
generally justified since both $\delta\sigma$ and $\delta\varphi$ are
stochastic quantities.

\section{evolution of isocurvature perturbation
after inflation}

In order to relate the spectrum of $\Phi$ at the end of inflation to
that at decoupling, we have to study the evolution of the isocurvature
mode together with the adiabatic mode after inflation.  
Unlike the adiabatic modes we do not have a universal formula for the
isocurvature modes, so we have to calculate them explicitly.

\subsection{isocurvature perturbation
in the reheating stage with Brans-Dicke dilaton}

First let us consider the isocurvature mode during the reheating stage 
after inflation. In this stage, $\kappa^2V \ll e^{-2a}V''$
(see Eq.(\ref{con3})), so that the inflaton oscillates
 coherently.  Since  equations of motion of perturbed
quantities, (\ref{vphi}) through (\ref{Phi}), have oscillating
coefficients, we must worry about possible parametric resonance
effect.  This issue has been studied by Kodama and Hamazaki \cite{kh}
recently in the case only the inflaton is present and oscillating.  
They have shown that 
resonance effect is unimportant at least in the long wavelength regime 
which is of our interest.  In the present case with two scalar fields, 
however,  their approach is not directly applicable, in particular, for 
the isocurvature mode \cite{yokoyama}.  Hence we have numerically solved
the evolution equations in the long wavelength limit.  
Figure 1 illustrates an example of the
results for the Brans-Dicke theory with $\omega=500$ and $\omega=5$.  
We find that there is no anomalous growth of perturbation variables 
and that linear analysis suffices even in this regime. 
We also find qualitatively the same results for more generic
scalar-tensor theories.

\subsection{Isocurvature perturbation
in the radiation-dominated stage with Brans-Dicke dilaton}

Next let us consider the isocurvature mode during the radiation
dominated era after the reheating stage. 
In appendix B, a general expression of the evolution equation 
for $\Phi$ is given in the dilaton-inflaton-radiation system. 
After the inflaton has decayed into radiation, 
Eq.(\ref{reheat}) becomes
\beqa
\ddot{\Phi}&+&(4+3c_s^2)H\dot{\Phi}+(2\dot{H}+3(1+c_s^2)H^2)\Phi
+c_s^2{k^2\over R^2}\Phi\nonumber\\
&=&{\kappa^2\over 3} 
{h_rh_d\over h}\lmk{\Delta_d\over 1+w_d}-{\Delta_r\over 1+w_r}\rmk \label{ee}
\eeqa
where $c_s$ is the sound velocity, 
\beq
c_s^2={\dot{p}\over \dot{\rho}}={3\dot{\vphi}^2 +4\rho_r/3 \over
3\dot{\vphi}^2 +4\rho_r},\label{cs2}
\eeq
with $h_j\equiv\rho_j + p_j$, $w_j\equiv p_j/\rho_j$ $(j=r, d)$, and 
$h\equiv\rho + p$.
It is now apparent that terms in the parenthesis in the right hand
side of (\ref{ee}) are just the entropy
perturbation
\beq
{\delta S\over S}={3\over 4}\Delta_r-{1\over 2}\Delta_d.
\eeq
For the isocurvature perturbation, it is nonvanishing. 
It is known that if we replace $\rho_d$ with the baryon density
$\rho_b$, the isocurvature mode grows as $R$\cite{hs}.
Because the dilaton density $\rho_d$ decays as $R^{-6}$ instead of 
$R^{-3}$, it is clear that the special solution originating from 
the source term decays as $R^{-2}$. 
On the other hand, the homogeneous part of the growing mode solution is 
the same as the adiabatic one.
Therefore the isocurvature mode in the radiation dominated stage 
grows no faster than the adiabatic mode.  
This behavior can be also seen by 
solving the Eq.(\ref{iso}) by using the Green's function method.  

To conclude, the isocurvature mode  does not evolve  
faster than the adiabatic counterpart and so is unimportant. 
The spectrum at the end of inflation is sufficient to compare the
observation.

\section{spectra of perturbations and constraints on PPN parameters}

Now we calculate the density fluctuation spectra produced by
inflation. In order to
do so, we have to specify $V(\sigma)$ and $a(\vphi)$. We consider 
chaotic inflation induced by a mass term  $V(\sigma)=m^2\sigma^2/2$,
of which we have a sensible particle-physics model in the intermediate 
scale \cite{MSYY}. 
Here $m$ is determined from
 the large-angle microwave anisotropies seen by COBE-DMR\cite{cobe}.

\subsection{general remark on initial conditions}

Initial conditions for $\vphi$ and $\sigma$ are given in terms of 
the number of e-foldings from the end of inflation;
\beq
N_I=-\int_{t_{e}} Hdt\simeq {\kappa\over 4}\int_{\vphi_{e}}{d\vphi\over
\alpha(\vphi)}\simeq \kappa^2\int_{\sigma_{e}}{e^{2a}V(\sigma)\over
V'(\sigma)}d\sigma , 
\eeq
where the suffix $e$ represents the end epoch of inflation.
Our present horizon ($H_0^{-1}=3000 h^{-1} {\rm Mpc}$) 
crossed outside the Hubble scale about 60
e-foldings before the end of inflation.
We assume that the slow-roll
condition is satisfied at $N_I$ e-folds from the end of inflation. 
Then the terms containing 
the two time derivative
can be neglected in Eqs.(\ref{vphi}-\ref{momp}). Initial conditions for 
$H,\dot{\vphi},\dot{\sigma}$ are given by
Eqs.(\ref{hams}-\ref{sigmas}). 

We need to determine $\vphi_{e}$ to calculate the spectrum of density 
perturbations. 
We normalize $\vphi$ in terms of its present value so that 
$\alpha^2(\vphi)= \alpha_0^2$ (see (\ref{ppn})) at present.  
Then we are able to specify  $\vphi_{e}$ by going back to the end of
inflation.

In order to connect with the present length scale, we have to estimate 
the number of e-foldings from the end of inflation to the present\cite{sbb}. 
After the end of inflation, 
the oscillating inflaton, which behaves like a dust fluid for the
massive inflaton,  dominates the universe, and then  radiation
dominated stage follows. 
The effective temperature, $T_{\rm rd}$, and the scale factor $R_{\rm
rd}$,  at the onset  of  radiation dominated
stage is determined
from the energy density
\beq
\rho_e\Bigl({R_e\over R_{\rm rd}}\Bigr)^3=g_{\rm eff}{\pi^2\over 30}T_{\rm rd}^4,
\eeq
where $\rho_e$ is the energy density at the end of inflation and 
$g_{\rm eff}$ is the effective number of degrees of freedom at
temperature $T$ typically  taking the value of $O(100)$ at $T_{\rm rd}$. 
Assuming the conservation of the entropy in relativistic particles per
comoving volume $s={4\over 3} R^3\rho/T$, we have
\beqa
R_{\rm rd}^3 g_{\rm eff}(T_{\rm rd})T_{\rm rd}^3&=&
(2T_{\gamma 0}^3+{21\over
4}T_{\nu 0}^3)R_0^3\nonumber\\
&=& {43\over 11} T_{\gamma 0}^3R_0^3,
\eeqa
where $T_{\nu 0}$ is the background neutrino temperature. 
The number of e-folding from the end of inflation to the present is then
\beqa
N_e&=&\ln{R_0\over R_e}\nonumber\\
& \simeq & 72+
{1\over 3}\ln{\rho_e\over m_{\rm pl}^4} -
{1\over 3}\ln{T_{\rm rd}\over m_{\rm pl}},
\eeqa
where $m_{\rm pl}$ is the Planck mass.
We set the present  value of the scale factor $R_0$ to unity for convenience. 

Given initial conditions, we solve Eqs.(\ref{back1}-\ref{back4}) 
numerically using a fourth-order
Runge-Kutta method. We choose $N_I=65$ typically. We express all
quantities in units of $m_{\rm pl}$ and use $x=\ln R$ for the integration
variable instead of the cosmic time\cite{sbb}.  

The calculations of the perturbed equations
Eqs.(\ref{vphi}-\ref{momp}) are begun when $k \leq RH$. 
The initial conditions for $\delta\vphi$ and 
$\delta\sigma$ are given by Eq.(\ref{init}).  Since the background
fields are in their slow-rolling phase, the initial conditions for
$\dot{\delta\vphi}, \dot{\delta\sigma}, \Phi$, and $\dot{\Phi}$ 
are given by neglecting terms with 
the second order time derivative in 
Eqs.(\ref{vphi}), (\ref{sigma}), (\ref{hamp}) and (\ref{momp}). 
We have to solve Eq.(\ref{adiab}) with
or without the source term for the evolution of $\Phi$ so that 
we can follow the evolution of adiabatic and
isocurvature modes separately. The spectral index $n$ defined by 
\beq
n-1={d \ln k^3\langle |\Phi|^2 \rangle\over d \ln k}
\eeq
is calculated for the present horizon scale.

\subsection{COBE normalization}

We have to normalize fluctuations to compare the spectra among
different gravitational theories. The COBE-DMR data give the
normalization. 
Following the standard treatment, 
the temperature fluctuations are expanded in terms of spherical harmonics as 
\beq
{\Delta T\over T}(\theta,\phi)=\sum_{lm}a_{lm}Y_{lm}(\theta,\phi),
\eeq 
and we define
\beq
\langle a_{lm}^*a_{l'm'}\rangle\equiv C_l \delta_{ll'}\delta_{mm'}.
\eeq
On COBE-DMR scales, temperature fluctuations are dominated by 
the Sachs-Wolfe effect which is expressed as 
$\Delta T(x)/T=\Phi(x)/3$\cite{sw} for 
scalar perturbations.  Here we neglect acoustic oscillation terms which 
may contribute to higher $l$'s.
Therefore $C_l$ is written as 
\beqa
C_l &\equiv& \langle |a_{lm}|^2 \rangle\nonumber\\
&=& {2\over \pi}\int dk k^2  \left\langle \left|{\Phi \over 3}
\right|^2\right\rangle j^2_l(k\eta_0),
\label{cl}
\eeqa
where $\eta_0 = 2H_0^{-1}$ is the conformal time at present.

If we approximate the power spectrum of $\Phi$ by a power law,
\beq
k^3\langle |\Phi|^2 \rangle =A\lmk {k\over H_0}\rmk ^{n-1},
\eeq
we can write Eq.(\ref{cl}) for scalar perturbations as
\beq
C_l= {A\over 36}{\Gamma(3-n)\Gamma(l+{n-1\over 2})\over
\Gamma^2({4-n\over 2})\Gamma(l+{5-n\over 2})}.
\eeq
The rms quadrupole $Q_{rms-PS}$ used by the COBE-DMR group is written in
terms of $C_2$ as  
\beq
Q_{rms-PS}= T_0\sqrt{5C_2\over 4\pi},
\eeq
where $T_0=2.726 {\rm K}$. According to the 4 yr COBE-DMR data \cite{cobe}
\beq
Q_{rms-PS}=18  \mu{\rm K},
\eeq
for $n=1$ Harrison-Zel'dovich spectrum.
It takes a slightly different value if 
different power low indices are considered, but we normalize the
spectra by using the above value because the index is varied around
$n=1$ and furthermore  the constraints
on the scalar-tensor theories are imposed primarily from the spectral
shape or its power-law index rather than from its amplitude.
The corresponding amplitude
$A$ is
\beq
A={144\pi\over5}{\Gamma^2({4-n\over 2})\Gamma(l+{5-n\over 2}) \over
\Gamma(3-n)\Gamma(l+{n-1\over 2})}\lmk {Q_{rms-PS}\over T_0}\rmk ^2.
\eeq
Eventually 
$m$ takes the 
value ranging from $O(10^{-6} m_{\rm pl})$ to $ O(10^{-8} m_{\rm pl})$.

\subsection{Brans-Dicke theory}

The Brans-Dicke theory corresponds to $\beta\equiv 0$ and thus 
\beqa 
\alpha(\vphi)&=&\alpha_0, \\
a(\vphi)&=&\alpha_0\kappa\vphi.
\eeqa
 Initial conditions for $\vphi$ and $\sigma$ are
\beqa
\kappa\vphi&=&4\alpha_0N+\kappa\vphi_{e},\\
\kappa\vphi_{e} &\simeq & 10\alpha_0,\\
\kappa^2\sigma^2&=&{2\over 3}e^{-2a(\vphi_e)}-{1\over 2\alpha_0^2}
(e^{-2a(\vphi)}-e^{-2a(\vphi_{e})}).
\eeqa

In Figs.\ 2, the spectra $k^3\langle |\Phi|^2 \rangle$ 
for each $\alpha_0$ or $\omega$ are
shown. In the Brans-Dicke theory, the conditions Eqs.(\ref{con1}) and
(\ref{con2}) hold and therefore we are justified to say that 
terms in proportion to $C_3$ represent the isocurvature mode. 
We see that considering the current limit on $\omega$ $(\omega > 
500)$ in Eq.(\ref{con5}), the contribution of the isocurvature mode is
totally negligible  in the Brans-Dicke theory in agreement with
Starobinsky and Yokoyama\cite{sy}. 

The spectral indices on the comoving horizon scale today are
$n=0.962, 0.966, 0.967$ for $\omega=500,5000,50000$, respectively. 
Note that $n=0.967$ in general relativity.

\subsection{a class of scalar-tensor theory}

We  take the following simple functional form for $a(\vphi)$ 
in accord with Damour and Nordtvedt\cite{dn}
\beqa
a(\vphi)&=&{\beta\over 2} {\kappa^2(\vphi^2 -
\vphi_0^2)}, ~~~~~\beta\equiv {\rm const,}\\
\alpha(\vphi)&=&\beta\kappa\vphi.
\eeqa
We choose the  origin of the field $\vphi$ so that 
$a(\vphi)= 0$ at present in order to reproduce the correct value of
the gravitational constant.\footnote{ 
The authors of \cite{bw} take $\alpha(\phi)=a_1+a_2\kappa\phi$
and choose the origin so that $\alpha=a_1$ at the end of inflation. 
However, $\alpha$ deviates from $a_1$ afterwards and may conflict with
the limit on $\alpha_0$.} 
We have to determine the value of $\vphi$ at the end of inflation,
{\it i.e.}, $\vphi_{e}$, to calculate the spectrum of density 
perturbation. Since we normalize $\vphi$ in terms of its present value, 
we need to go back to the end of inflation to specify $\vphi_{e}$. 
The evolution of $\vphi$ in the radiation or matter-dominated universe 
is analyzed by Damour and Nordtvedt\cite{dn}(see also \cite{bm}). 
In the radiation-dominated stage $\vphi$ is hardly changed. We assume
for simplicity that $\vphi_{e}$ is equal to $\vphi$ at the beginning
of matter dominated stage. For $|\beta| \ll 1$(see Eq.(\ref{con5})), 
 $\vphi_{e}$ is determined by 
\beq
\kappa\vphi_{e}\simeq{\alpha_0\over \beta}e^{10\beta}.
\eeq
Then $\vphi$ is given in terms of $N$ as
\beq
\kappa\vphi\simeq \kappa\vphi_{e}e^{4\beta N}\simeq {\alpha_0\over
\beta}e^{4\beta N+10\beta}.
\label{initdil}
\eeq
For $\sigma$ with (see Eq.(\ref{con2}))
\beq
e^{-2a(\vphi_{e})}
\lmk {V'(\sigma_{e})\over \kappa V(\sigma_{e})}\rmk ^2=6,
\eeq
$\sigma$ is given by
\beq
\kappa^2\sigma^2\simeq {2\over 3}e^{-2a(\vphi_{e})}
+{e^{\alpha_0^2/\beta}\over 2\beta}\lkk 
E_i(-\beta(\kappa\vphi)^2 )-E_i(-\beta(\kappa\vphi_e)^2)
\rkk ,
\eeq
where $E_i(x)$ is the exponential integral function defined by 
$E_i(-x)=-\int^{\infty}_xe^{-t}dt/t$.

In Figs.\ 3, the spectra for several $\beta$ with $\alpha_0=10^{-3}$ are
shown. In some cases the mode coming from $C_3$ part in 
Eq.(\ref{c3}) is not negligible and the spectrum can be different
from flat one. 
In  scalar-tensor theories with positive $\beta$, the conditions
Eqs.(\ref{cond1}) and (\ref{cond2}) do not hold 
since initially $\alpha$ can be greatly different from zero (see
Eq.(\ref{initdil})), and terms in proportion to $C_3$ contain 
the adiabatic mode in addition to the isocurvature mode. 
For  theories with negative $\beta$, however, 
terms in proportion to $C_3$ represent the isocurvature mode. 

In Fig.\ 4 we show that contour 
plot of $n$ on $\alpha_0-\beta$ plane. We see that for $\beta>0$ $n$
falls down very steeply with  increasing  $\beta$. 
Thus the observed spectrum by COBE-DMR\cite{cobe}, $n=1.2\pm 0.3$, 
strongly constrains $\beta$. 
This possibility was pointed out in \cite{bw} for  $\beta>0$. 
Interestingly, the contour level can be
fitted by a linear function in $\log \alpha_0$. For $n \geq 0.9$, 
\beq
9.0\times 10^{-3} \log \lmk {\alpha_0\over \sqrt{5}\times 10^{-2}}\rmk 
 -1.5\times 10^{-2} \leq \beta \leq 
-8.9\times 10^{-3} \log \lmk {\alpha_0\over \sqrt{5}\times 10^{-2}}\rmk 
 +5.3\times 10^{-3}. \label{const1}
\eeq
For $n \geq 0.7$,
\beq
5.0\times 10^{-3} \log \lmk {\alpha_0\over \sqrt{5}\times 10^{-2}}\rmk 
-3.7\times 10^{-2} \leq \beta \leq 
-9.2\times 10^{-3} \log \lmk {\alpha_0\over \sqrt{5}\times 10^{-2}}\rmk 
 +7.3\times 10^{-3}. \label{const2}
\eeq
Since the tensor mode of perturbations does not contribute 
to the spectrum significantly in the above range of the spectral
index, the above inequality gives the constraint on $\beta$.  
Note  that the constraint on $\beta$ has 
little dependence on $\alpha_0$ as well as on the details of
normalization of fluctuations; adopting the inflaton mass which is ten 
times larger results in only a few percent changes in the constraint. 
We  also note that the constraint is stronger by
an order-of-magnitude than those obtained by post-Newtonian
experiments(see Eq.(\ref{ppn2})).

\section{summary}

We have studied  density perturbations 
produced during chaotic inflation taking both adiabatic and
isocurvature modes into account in a class of  scalar-tensor theories
of gravity in which the dilaton coupling is metric one.  
The spectrum of the density perturbation produced 
by chaotic inflation in a scalar-tensor theory 
can be a non-flat one because of the variable Brans-Dicke coupling. 
The spectrum observed by COBE-DMR thus constrains such a coupling. 
Assuming a simple coupling function, the constraint is found to be
stronger by an order-of-magnitude than those obtained by post-Newtonian
experiments. 
It is interesting to note that in the case of the simple coupling
function $\alpha(\vphi)=\beta\vphi$ employed here, our constraint is
much more stringent than that obtained by the binary pulsar
experiment\cite{de2}, $\beta > -5$, in a sense.  Indeed $\alpha_0$ must 
be $\alpha_0 < 10^{-1000}$(!) so that the constraint Eq.(\ref{const1}) or 
Eq.(\ref{const2}) is looser than $\beta > -5$ if these fitting
formulas are extrapolated thus far. 
Thus we may conclude that cosmological inflation is suited in the
framework of  general relativity, or more advocatively, inflation favors
Einstein.

\acknowledgments

The authors are grateful to J.D.Barrow and the anonymous referee 
for useful comments.
T.C. would like to thank T.Nakamura and H.Sato for continuous
encouragement.
Numerical calculations were supported by Yukawa Institute for
Theoretical Physics.
This work was partially supported by the Japanese Grant
in Aid for Science Research Fund of the Ministry of Education,
Science, Sports
and Culture Nos.\ 07304033(JY) and 08740202(JY).

\appendix
\section{FORMULAE FOR ADIABATIC MODE}
\label{sc:formulae}

In this appendix we review derivation of the formulae for the adiabatic mode
Eqs.(\ref{ad1}) and (\ref{ad2}). See \cite{ps} for a different approach.

Using Eqs.(\ref{Phi}), (\ref{hamp}) and (\ref{momp}), we get
\beqa
\ddot{\Phi}+(4+3c_s^2)H\dot{\Phi}&+&(2\dot{H}+3(1+c_s^2)H^2)\Phi
+c_s^2{k^2\over R^2}\Phi\nonumber\\
&=&(c_s^2-1){k^2\over R^2}\Phi+\kappa^2{e^{4a}\over
\dot{\vphi}^2+e^{2a}\dot{\sigma}^2}(V'\dot{\vphi}-4e^{2a}\alpha\kappa
V\dot{\sigma})(\dot{\sigma}\delta\vphi-\dot{\vphi}\delta\sigma),\label{adiab}
\eeqa
where $c_s^2$ is defined by
\beq
c_s^2={\dot{p}\over \dot{\rho}}=1+{2e^{4a}(4\alpha\kappa
V\dot{\vphi}+V'\dot{\sigma}) \over
3H(\dot{\vphi}^2+e^{2a}\dot{\sigma}^2)}.
\eeq
The left-hand-side of Eq.(\ref{adiab}) is just the same as that of the 
adiabatic perturbation for hydrodynamical matter and in case of 
the adiabatic perturbation right-hand-side term vanishes. 
For long wavelength perturbations right-hand-side vanishes when
\beq
{\delta\vphi\over \dot{\vphi}}={\delta\sigma\over \dot{\sigma}}.
\eeq
To derive the solution for the adiabatic mode Eq.(\ref{ad1}), it is
convenient to use the constancy of the Bardeen's $\zeta$
\beq
\zeta \equiv -{H^2\over \dot{H}}(\Phi+H^{-1}\dot{\Phi})+\Phi ={\rm
const}\equiv C_1,
\eeq
which follows from the fact that  
the left-hand-side of Eq.(\ref{adiab}) is equal 
to $-\dot{H}\dot{\zeta}/H$ in the long-wavelength limit. 
Substituting $\Phi-C_1={\tilde \Phi}$, the above equation becomes a
homogeneous form
\beq
\dot{\tilde \Phi}+H{\tilde \Phi}-{\dot{H}\over H}{\tilde \Phi}
=-C_1H.
\eeq
Noting the left-hand-side is written as 
\beq
{H\over R}\lmk {R\over H}{\tilde \Phi}\rmk ^{.},
\eeq
the solution is immediately given as
\beq
\tilde{\Phi}=-C_1{H\over R}\int R(t')dt'+C_2{H\over R},
\eeq
where $C_2$ corresponds to the amplitude of the decaying mode which is 
neglected in Eq.(\ref{ad1}).

To derive Eq.(\ref{ad2}), we use Eq.(\ref{momp}) as 
\beqa
\dot{\Phi}+H\Phi&=&-C_1{\dot{H}\over R}\int Rdt' +
C_2{\dot{H}\over R}\nonumber\\
&=&{\kappa^2\over
2}(\dot{\vphi}\delta\vphi+e^{2a}\dot{\sigma}\delta\sigma)
={\kappa^2\over 2}{\delta\vphi\over \dot{\vphi}}(\dot{\vphi}^2+e^{2a}
\dot{\sigma}^2).
\eeqa
Employing Eq.(\ref{back4}) we  find 
\beq
{\delta\vphi\over \dot{\vphi}}={\delta\sigma\over \dot{\sigma}}=
{1\over R}\lmk C_1\int Rdt'-C_2\rmk .
\eeq
Neglecting the decaying mode, we get Eq.(\ref{ad2}).

\section{LINEAR PERTURBATIONS FOR DILATON-INFLATON-RADIATION SYSTEM}
\label{sc:linear}

Basic equations in the Einstein frame are the Einstein equation
\beq
G_{ab}=\kappa^2\lmk T_{ab}+\p_a\vphi\p_b\vphi
-{1\over2}g_{ab}g^{cd}\p_c\vphi \p_d\vphi
+e^{2a}\p_a\sigma\p_b\sigma
-g_{ab}({1\over 2}g^{cd}e^{2a}\p_c\sigma \p_d\sigma
-e^{4a}V(\sigma))\rmk ,
\eeq
the dilaton equation of motion
\beq
\Box \vphi=-\kappa \alpha(\vphi) (T-g^{cd}e^{2a}\p_c\sigma \p_d\sigma
+4e^{4a}V(\sigma)),
\eeq
the matter equation of motion (Bianchi identity)
\beq
\nabla _b{T^b}_a=\kappa \alpha (T-g^{cd}e^{2a}\p_c\sigma \p_d\sigma
+4e^{4a}V(\sigma)) \nabla _a\vphi,
\eeq
and the inflaton equation of motion
\beq
\Box\sigma=e^{2a}V'(\sigma)-2\kappa\alpha g^{ab}\vphi_{,a}\sigma_{,b},
\eeq
where $\kappa^2=8\pi G$.

\subsection{background equations}

The matter energy momentum tensor is given by
\beq
T_{ab}=\rho_r u_au_b+p_r(g_{ab}+u_au_b).
\eeq
Assuming the flat FRW model, the background equations are
\beqa
H^2&=&{\kappa^2\over 3}\lmk \rho_r +{1\over 2}\dot{\vphi}^2+
{1\over 2}e^{2a}\dot{\sigma}^2+e^{4a}V(\sigma)\rmk ,\\
\dot{H}&=&-{\kappa^2\over 2}\lmk \rho_r+p_r+
\dot{\vphi}^2+e^{2a}\dot{\sigma}^2\rmk ,\\
\ddot{\vphi}+3H\dot{\vphi}&=&\kappa
\alpha(-\rho_r+3p_r+e^{2a}\dot{\sigma}^2
-4e^{4a}V(\sigma)),\\
(\rho_rR^3)^{\dot{}}+p_r(R^3)^{\dot{}}
&=&(\rho_r-3p_r)R^3\dot{a}(\vphi),\\
\ddot{\sigma}+3H\dot{\sigma}&=&-e^{2a}V'(\sigma)-
2\kappa\alpha\dot{\vphi}\dot{\sigma},
\eeqa
where $p_r=\rho_r/3$. 

\subsection{linear perturbations}

We employ the metric perturbation in the longitudinal gauge
as  
\beq
ds^2=-(1+2\Phi)dt^2+R^2(1-2\Phi)d{\bf x}^2.
\eeq
Then the perturbed
Einstein tensors  are\cite{ks}
\beqa
\delta {G^t}_t&=&{2\over R^2}\lkk 3\dot{R}^2\Phi +3R\dot{R}\dot{\Phi}
+k^2\Phi\rkk ,\\
\delta {G^t}_j&=&{2k\over R^2}\lkk \dot{R}\Phi +R\dot{\Phi}\rkk Y_j,\\
\delta {G^i}_j&=&{2\over
R^2}\lkk (\dot{R}^2+2R\ddot{R})\Phi+R\dot{R}\dot{\Phi}
+(R^2\dot{\Phi})^{.} +R\dot{R}\dot{\Phi}\rkk \delta^i_j.
\eeqa
As for the matter components, hydrodynamical perturbations of radiation 
are
\beqa
\delta {T^t}_t&=&-\rho_r\delta,\label{delgamma}\\
\delta {T^t}_j&=& (\rho_r+p_r)vY_j,\\
\delta {T^i}_j&=& p_r\pi_L\delta^i_j.
\eeqa
The dilaton perturbations are expressed as 
\beqa
\delta {T^t}_t&=& \dot{\vphi}^2\Phi-\dot{\vphi}\dot{\delta\vphi},
\label{deldila}\\
\delta {T^t}_j&=& {k\over R}\dot{\vphi}\delta\vphi Y_j,\\
\delta {T^i}_j&=& \lkk -\dot{\vphi}^2\Phi
+\dot{\vphi}\dot{\delta\vphi}\rkk \delta^i_j.
\eeqa
and the inflaton perturbations as
\beqa
\delta {T^t}_t&=&
-e^{2a}(\dot{\sigma}\dot{\delta\sigma}-\dot{\sigma}^2\Phi)
-\kappa\alpha e^{2a}\dot{\sigma}^2\delta\vphi,
\label{delinf}\\
\delta {T^t}_j&=& {k\over R}e^{2a}\dot{\sigma}\delta\sigma Y_j,\\
\delta {T^i}_j&=&
\lkk e^{2a}(\dot{\sigma}\dot{\delta\sigma}-\dot{\vphi}^2\Phi)
+\kappa\alpha e^{2a}\dot{\sigma}^2\delta\vphi\right. \nonumber\\
&-&\left. e^{4a}V'(\sigma)\delta\sigma
-4\kappa\alpha e^{4a}V(\sigma)\delta\vphi
\rkk \delta^i_j.
\eeqa

Thus we have the perturbed Einstein equations
\beqa
2\lkk 3\dot{H}\dot{\Phi}+ 3H^2\Phi
+{k^2\over R^2}\Phi\rkk &=&-\kappa^2\lkk 
-(\dot{\vphi}^2+e^{2a}\dot{\sigma}^2)\Phi
+\rho_r\delta+\dot{\vphi}\dot{\delta\vphi}+
e^{2a}\dot{\sigma}\dot{\delta\sigma}\right.\nonumber\\
&+&\left.\kappa\alpha e^{2a}\dot{\sigma}^2\delta\vphi
+e^{4a}V'(\sigma)\delta\sigma+
4\kappa\alpha e^{4a}V(\sigma)\delta\vphi
\rkk ,\\
2\lkk \dot{\Phi}+H\Phi\rkk &=& \kappa^2\lkk (\rho_r+p_r)v{R\over k} 
+\dot{\vphi}\delta\vphi +e^{2a}\dot{\sigma}\delta\sigma\rkk ,\\
2\lkk \ddot{\Phi}+4H\dot{\Phi}+(2\dot{H}+3H^2)\Phi\rkk &=&
\kappa^2\lkk 
-(\dot{\vphi}^2+e^{2a}\dot{\sigma}^2)\Phi
+{1\over 3}\rho_r\delta+\dot{\vphi}\dot{\delta\vphi}+
e^{2a}\dot{\sigma}\dot{\delta\sigma}\right.\nonumber\\
&+&\left.\kappa\alpha e^{2a}\dot{\sigma}^2\delta\vphi
-e^{4a}V'(\sigma)\delta\sigma-
4\kappa\alpha e^{4a}V(\sigma)\delta\vphi
\rkk .
\eeqa

The perturbed equations are rewritten as
\beqa
\ddot{\Phi}+(4+3c_s^2)H\dot{\Phi}&+&(2\dot{H}+3(1+c_s^2)H^2)\Phi
+c_s^2{k^2\over R^2}\Phi\nonumber\\
=(c_s^2-1){k^2\over R^2}\Phi&-&{\kappa^2\over 2}\lkk 
{2\over 3}\rho_r\delta +2e^{4a}V'(\sigma)\delta\sigma +8\kappa\alpha
e^{4a}V(\sigma)\delta\vphi\right.\nonumber\\
&+&\left.{1\over h}(2Hh_r-2^{4a}V'(\sigma)\dot{\sigma}-8\kappa\alpha
e^{4a}V(\sigma)\dot{\vphi})(h_r{vR\over  k}
+\dot{\vphi}\delta\vphi+e^{2a}\dot{\sigma}\delta\sigma)
\rkk ,
\label{iso}
\eeqa
where $c_s^2$ is defined by
\beq
c_s^2={\dot{p}\over \dot{\rho}}={4H\rho_r/3+
3H(\dot{\vphi}^2+e^{2a}\dot{\sigma}^2) +2e^{4a}V'(\sigma)\delta\vphi+
8\kappa\alpha e^{4a}V(\sigma)\delta\vphi
\over
4H\rho_r+3H(\dot{\vphi}^2+e^{2a}\dot{\sigma}^2)}.\label{cs}
\eeq
We have arranged the left-hand-side of Eq.(\ref{iso}) so that  
it is just the same as that of the 
adiabatic perturbation for hydrodynamical matter. 

Now let us examine the structure of the right-hand-side of Eq.(\ref{iso}) 
in detail. Before doing so, we define the following useful gauge
invariant variables\cite{ks}. For a general multi-component system 
$(\rho_a,p_a)$, $\delta_a$ and $v_a$ are defined by 
\beqa
\delta {T^t}_t&=&-\rho_a\delta_a,\\
\delta {T^t}_j&=&h_av_aY_j,
\eeqa
where $h_a=\rho_a+p_a$. The associated gauge invariant variables
$\Delta_a$ and $V_a$ are defined by
\beqa
\Delta_a&=&\delta_a+3(1+w_a)(1-q_a){RH\over k}V_T,\\
V_a&=&v_a,
\eeqa
where $w_a=p_a/\rho_a$ and $V_T=\Sigma h_aV_a/h$ with $h=\Sigma h_a$ 
and $q_a$ is the source term for each energy momentum tensor.
For the case we are considering, i.e., the radiation-inflaton-dilaton system, 
the respective $\Delta_a(\Delta_r,\Delta_i,\Delta_d)$ are written as
\beqa
{\Delta_r\over 1+w_r}&=& {3\over 4} \delta +3{RH\over
  k}\lkk 1+{(e^{2a}\dot{\sigma}^2-4e^{4a}V(\sigma))\kappa\alpha\dot{\vphi}
\over 3Hh_r}\rkk V_T,\\
{\Delta_i\over 1+w_i}&=& \Phi +{\dot{\delta\sigma} \over \dot{\sigma}}
+\kappa\alpha\delta\vphi +3{RH\over k}V_T,\\
{\Delta_d\over 1+w_d}&=&\Phi +{\dot{\delta\vphi} \over \dot{\vphi}}
+3{RH\over k}
\lkk 1-{(e^{2a}\dot{\sigma}^2-4e^{4a}V(\sigma))\kappa\alpha\dot{\vphi}
\over 3Hh_d}\rkk V_T.
\eeqa

Then Eq.(\ref{iso}) can be rewritten as  
\beqa
\ddot{\Phi}+(4+3c_s^2)H\dot{\Phi}&+&(2\dot{H}+3(1+c_s^2)H^2)\Phi
+c_s^2{k^2\over R^2}\Phi\nonumber\\
=(c_s^2-1){k^2\over R^2}\Phi&-&{\kappa^2\over 2}\lkk 
{1\over 2}h_r\delta +2h_rV_T{RH\over k}\right.\nonumber\\ 
&+&2{h_r\over h}e^{2a}V'(\sigma)\dot{\sigma}\lmk {\delta\sigma\over
  \dot{\sigma}}-{vR\over k}\rmk 
+8{h_r\over h}\kappa\alpha
e^{4a}V(\sigma)\dot{\vphi}\lmk {\delta\vphi\over \dot{\vphi}}-{vR\over
    k}\rmk \nonumber\\
&+&\left.{2e^{4a}\dot{\sigma}\dot{\vphi}\over h}(V'(\sigma)\dot{\vphi}-
4\kappa\alpha e^{2a}V(\sigma)\dot{\sigma})\lmk {\delta\sigma\over
  \dot{\sigma}}-{\delta\vphi\over \dot{\vphi}}\rmk \rkk .
\eeqa
When radiation is absent the expression agrees with that of inflaton-dilaton system.

By expanding out the term $(c_s^2-1){k^2\over R^2}\Phi$, we finally have
\beqa
\ddot{\Phi}&+&(4+3c_s^2)H\dot{\Phi}+(2\dot{H}+3(1+c_s^2)H^2)\Phi
+c_s^2{k^2\over R^2}\Phi\nonumber\\*
&=&-{\kappa^2\over 2}\lnk {2\over 3h}\lkk  h_rh_i\lmk{\Delta_r\over 1+w_r}-
{\Delta_i\over 1+w_i}\rmk+h_rh_d\lmk{\Delta_r\over 1+w_r}-
{\Delta_d\over 1+w_d}\rmk\rkk -
{2RV_T\over 3k}(\dot{\sigma}^2-4V(\sigma))\kappa\alpha\dot{\vphi}\right.
\nonumber\\*
&+&{1\over 3Hh}(2V'(\sigma)e^{4a}\dot{\sigma}+8\kappa\alpha
e^{4a}V(\sigma)\dot{\vphi})(\rho\Delta+e^{2a}V'(\sigma)\delta\sigma+
4\kappa\alpha e^{4a}V(\sigma)\delta\vphi)\nonumber\\*
&+&2{h_r\over h}e^{2a}V'(\sigma)\dot{\sigma}\lmk {\delta\sigma\over
  3\dot{\sigma}}-{vR\over k}\rmk 
+8{h_r\over h}\kappa\alpha
e^{4a}V(\sigma)\dot{\vphi}\lmk {\delta\vphi\over 3\dot{\vphi}}-{vR\over
    k}\rmk \nonumber\\*
&+&\left.{2e^{4a}\dot{\sigma}\dot{\vphi}\over h}(V'(\sigma)\dot{\vphi}-
4\kappa\alpha e^{2a}V(\sigma)\dot{\sigma})\lmk {\delta\sigma\over
  \dot{\sigma}}-{\delta\vphi\over \dot{\vphi}}\rmk \rnk .\label{reheat}
\eeqa

\newpage
\vskip 0.3in
\centerline{FIGURE CAPTION}
\vskip 0.05in

\newcounter{fignum}
\begin{list}{Fig.\arabic{fignum}.}{\usecounter{fignum}}

\item
Evolution  of the inflaton and the curvature perturbations during the 
oscillation regime of the inflaton in the Brans-Dicke theory 
with (a)$\omega=500$ and (b)$\omega=5$.  
The abscissa is the e-folding number after the end of inflation.

\item
The spectra of perturbations in Brans-Dicke theory with
$\omega=500,5000,50000$. The abscissa is
the wavenumber in unit of $h {\rm Mpc}^{-1}$. 
Logarithm is base 10. 
The isocurvature mode is negligible in this theory. 

\item 
The spectra of perturbations in scalar-tensor theories with
$\alpha_0=1.0\times 10^{-3}$ and $\beta=0.02,0.005,-0.04$. 
For larger $|\beta|$, the mode coming from $C_3$ part in 
Eq.(\ref{c3}) is not negligible  and 
the spectrum has slope. The respective spectral index is 
$n=0.640,0.967,0.734$.

\item 
The contour plot of spectral index $n$ for scalar-tensor theories. 
The contour levels $n=0.9,0.7$ are shown. The region bounded by these
courves is the allowed region. We find non-zero $\beta $
theories are strongly constrained. The shaded region is excluded by 
the solar system experiments; the vertical line is the constraint by 
the expetiment of the deflection of light (or the time delay of
light): $\alpha_0 < \sqrt{5}\times 10^{-2}$.
The dotted curve in the shaded region is the constraint by the
lunar-raser-ranging experiment: $\alpha_0^2|\beta_0| < 3\times 10^{-4}$.

\end{list}

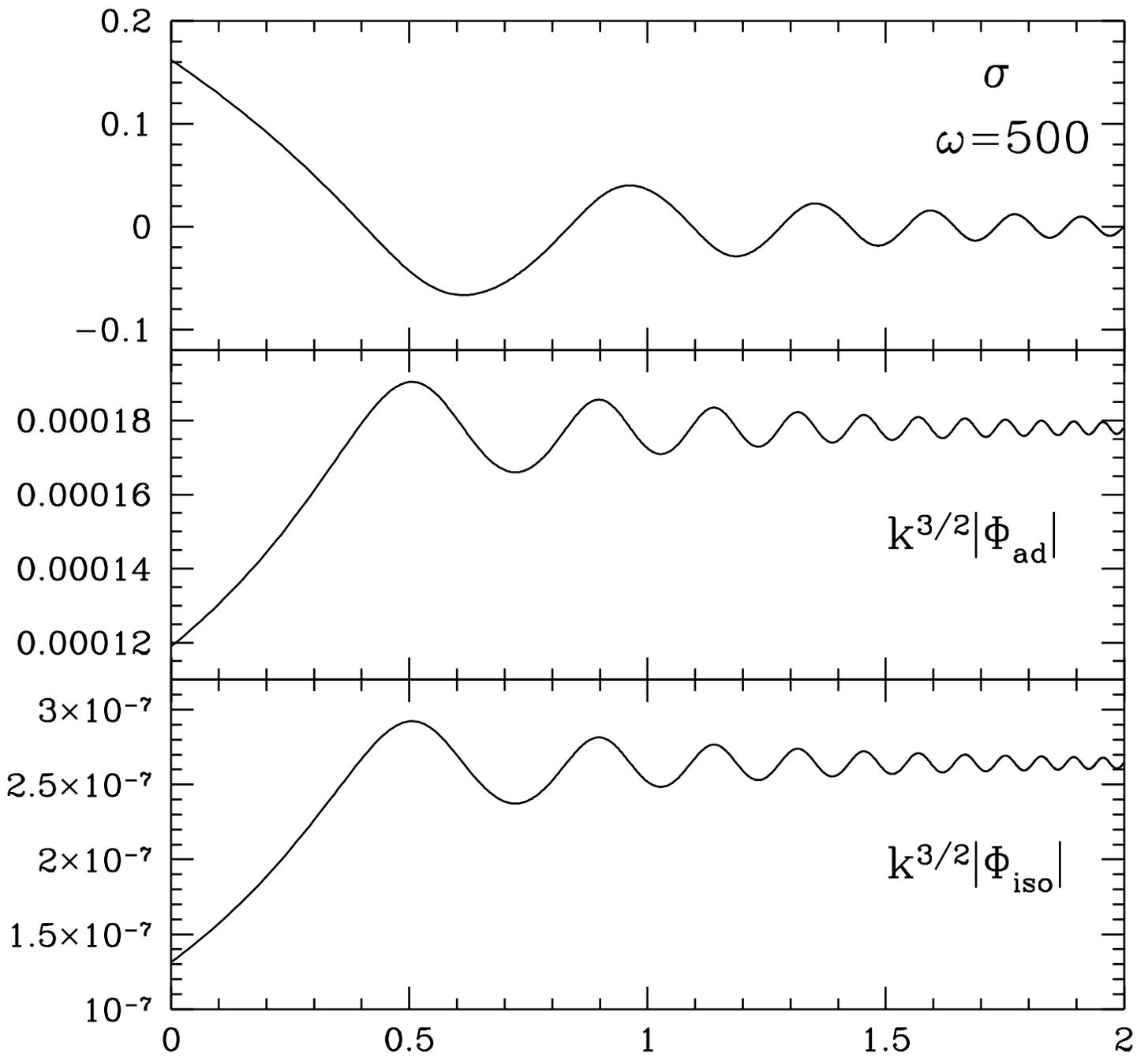
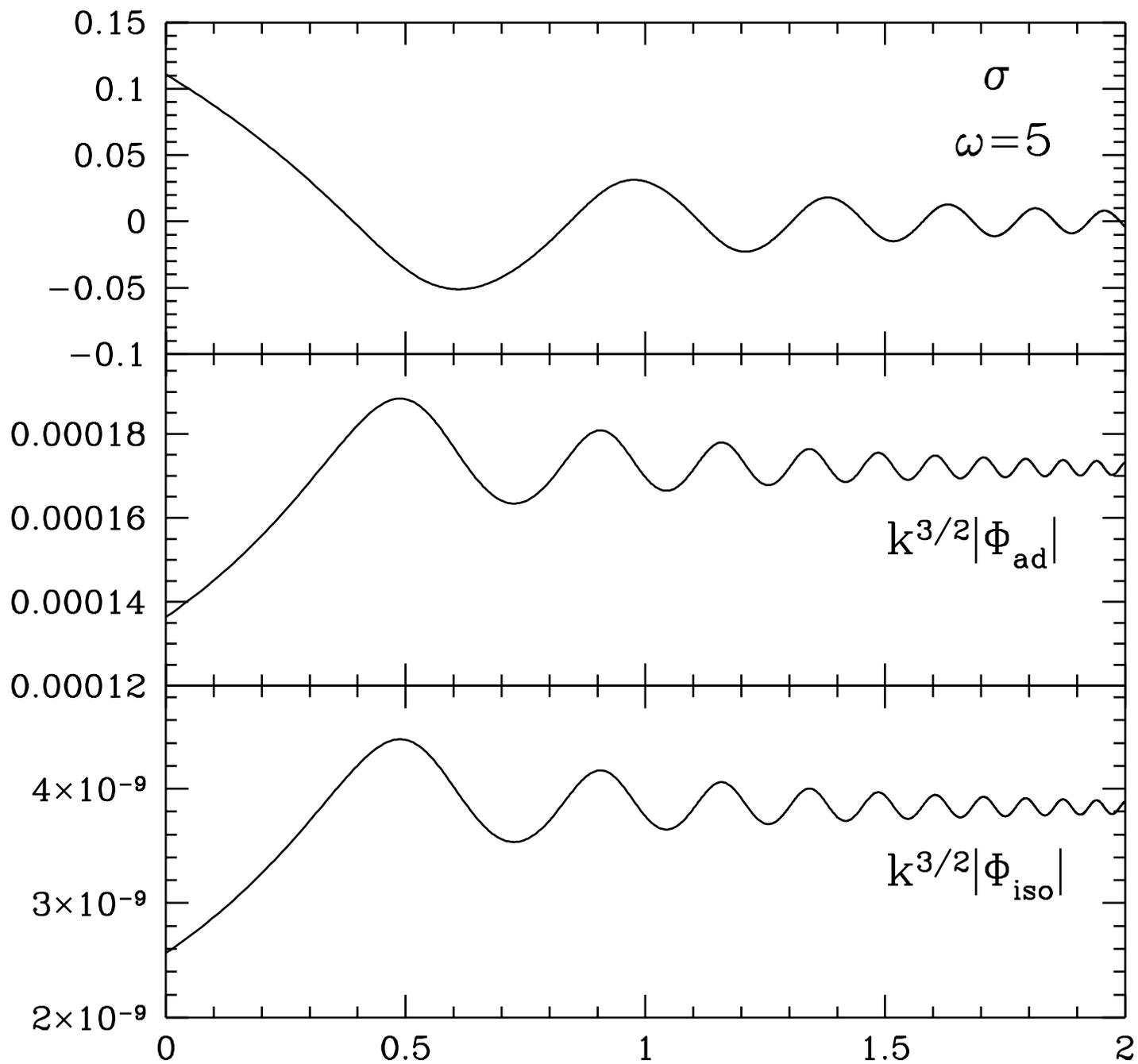
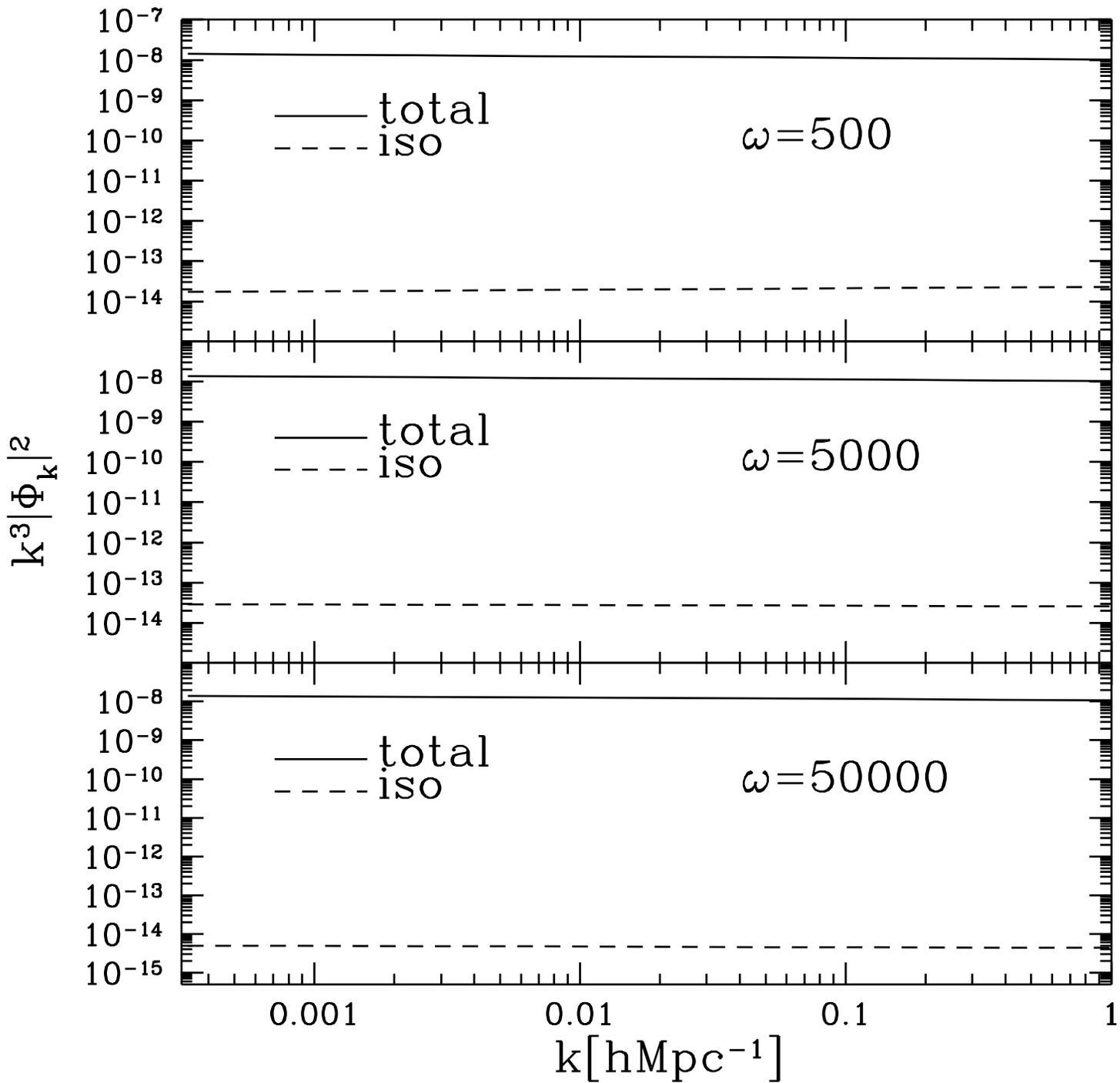
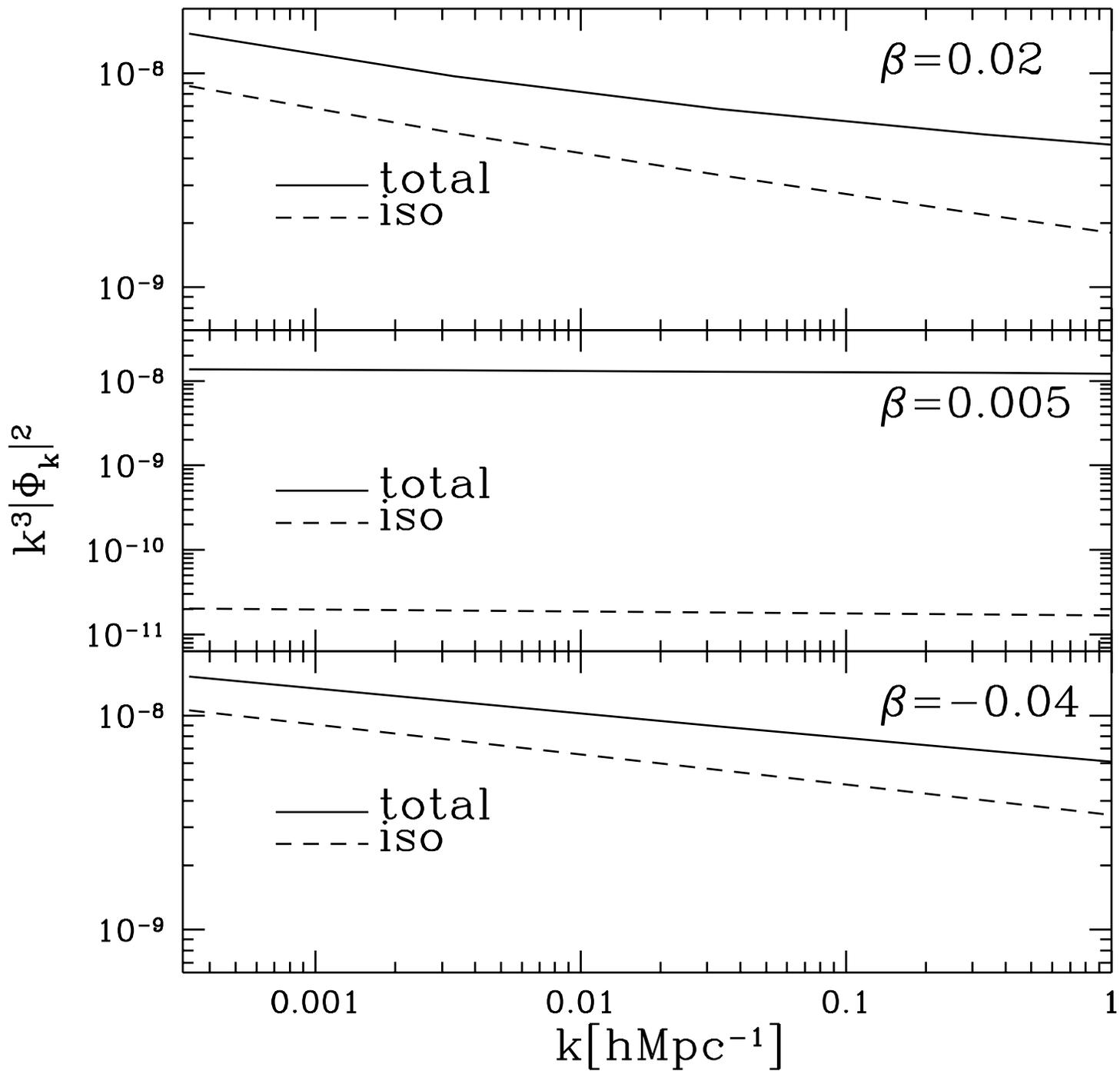
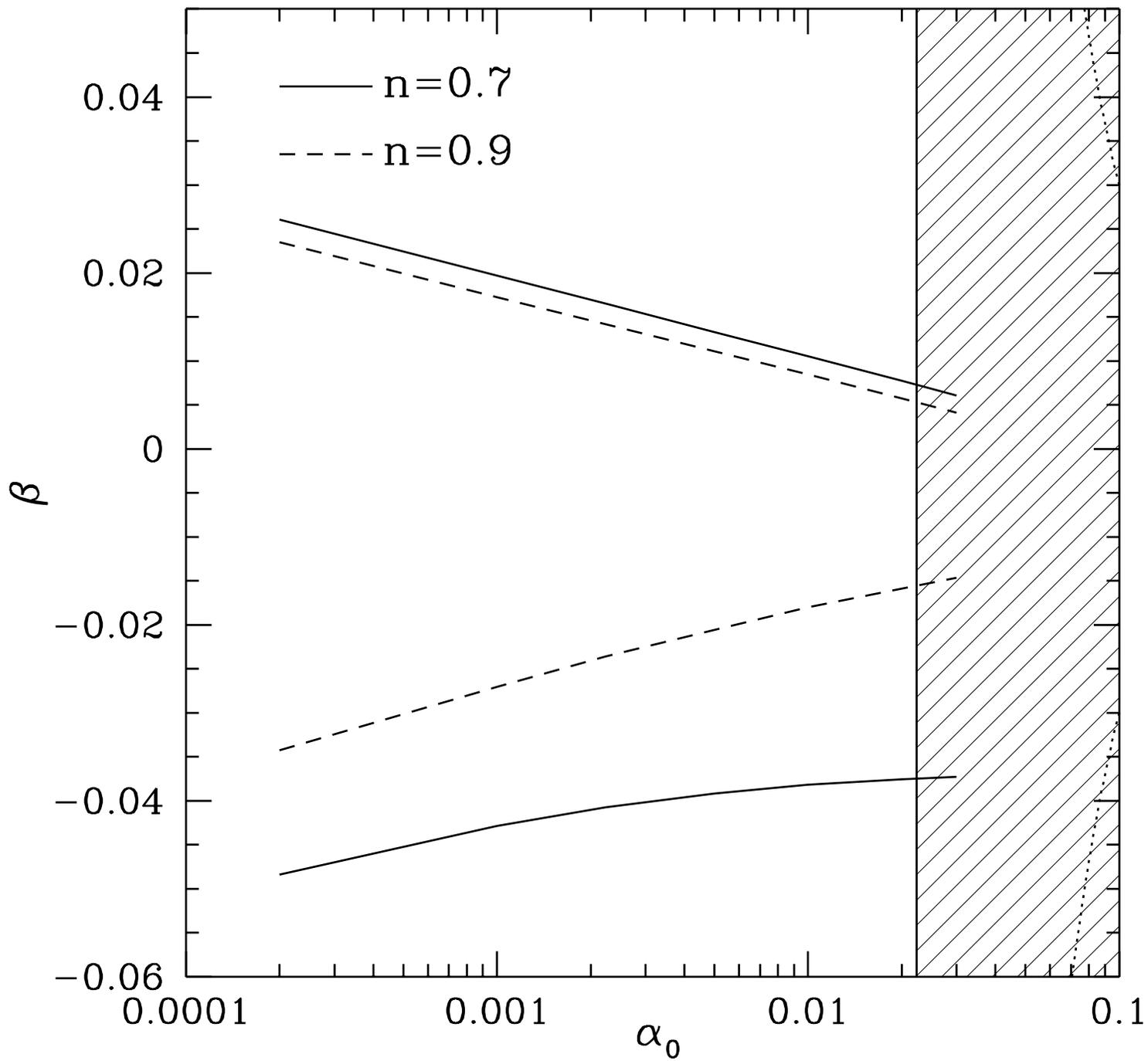
\end{document}